\documentstyle[amsmath,amssymb,graphicx]{article}

\def\be{\begin{eqnarray}}
\def\ee{\end{eqnarray}}
\def\nn{\nonumber}

\def\tr{{\rm tr}\,}

\hoffset= - 1.0in         % switch off for draft style
\begin{document}

\hfill ITEP/TH-24/09

\bigskip

\centerline{\Large{On Equivalence of two Hurwitz Matrix Models
}}

\bigskip
\begin{center}
%\vspace{0.3cm}
%\emph{} }\\
{\large A.Morozov and Sh.Shakirov}\\
\vspace{0.4cm}
{\em ITEP, Moscow, Russia\\
MIPT, Dolgoprudny, Russia}\\
\end{center}

\bigskip

\centerline{ABSTRACT}

\bigskip

{\footnotesize
In
%arXiv:0902.2627
\cite{Wops}
a matrix model representation was found
for the simplest Hurwitz partition function,
which has Lambert curve $\phi e^{-\phi} = \psi$
as a classical equation of motion.
We demonstrate that Fourier-Laplace transform
in the logarithm of external field $\Psi$
converts it into a more sophisticated form,
recently suggested in
%arXiv:0906.1206.
\cite{BEMS}.
}

\bigskip

\bigskip

Representation of Riemann surfaces (curves) as ramified coverings
of some simpler (bare) surfaces, like Riemann sphere,
is a very important technique in string theory,
used from summation of perturbation theory series \textit{a la}
\cite{KniUFN,UFN2} to description of partition functions
as $\tau$-functions of integrable hierarchies \cite{intau}-\cite{DV}.
For many years, however, this subject looked too difficult to
be addressed in a systematic way, and only recently the
underlying theory of Hurwitz-Kontsevich (HK) partition functions
-- concise language for accumulating information about
ramified coverings -- began to attract attention.
Fortunately, now activity in this field is fastly developing,
see \cite{Hur} for some important references.
This theory is only in its very beginning and it is, perhaps,
too early to apply it to the above-mentioned old problems
of general value
(of which the main one is development of the free-field calculus
{\it a la} \cite{KniUFN,fref} on arbitrary coverings),
but it is quite interesting and puzzling by itself.
Being related to explicitly topological theories,
the HK partition functions should of course possess matrix-model
representations, and their place in the "M-theory of matrix models"
\cite{Mth} is one of the first problems to be addressed in this area.

Recently, in the simplest example of "ordinary" Hurwitz numbers
such matrix model was explicitly derived in \cite{Wops}.
It belongs to the family of Generalized Kontsevich models
\cite{GKM,UFN3}, but has a somewhat unusual form:

\begin{equation}
\addtolength{\fboxsep}{5pt}
\boxed{
\begin{gathered}
Z_{HK}(\Psi) \sim \int_{n\times n}
d \mu(\phi) \
\exp\left( - \dfrac{1}{2g} \tr \phi^2 +
\tr \big( e^{\phi - ng/2} \Psi \big) \right) \end{gathered}
}\label{HarerZagierDuo}
\end{equation}
\smallskip\\
where $\Psi$ is an $n\times n$ matrix-valued Miwa variable, through which
the ordinary time-variables are expressed as
$p_k = kt_k = \tr \Psi^k$, and the integral is taken over all Hermitian $n \times n$ matrices $\phi$ with the measure

\begin{align}
d\mu(\phi) = \sqrt{{\det} \left(
\dfrac{ \sinh \left(\frac{\phi\otimes I - I\otimes\phi}{2}\right)}
{ \left(\frac{\phi\otimes I - I\otimes\phi}{2}\right)} \right)} d\phi
\ =\  \Delta(\phi)\Delta(e^{\phi})\
[dU] \prod\limits_{i=1}^n e^{-\frac{n-1}{2} \phi_i} d\phi_i
\label{meas}
\end{align}
\smallskip\\
This measure is one of the main peculiarities of the model.
It is in fact nicely matched with the exponential coupling to
external field $\Psi$ in the action, so that trigonometric
Van-der-Monde determinant
$\Delta(e^\phi) = \prod_{i<j} (e^{\phi_i}-e^{\phi_j})$
is absorbed into the Itzykson-Zuber integral over unitary matrices $U$.
Another attractive feature is that in neglect of this
measure the classical equation of motion has the desired
form of Lambert curve,
\begin{align}
\phi e^{-\phi} = e^{-ng/2}\psi,
\end{align}
which is well known from many papers of ref.\cite{Hur}
to play an important role in the theory of Hurwitz numbers.
Last week an alternative matrix model
representation was suggested in \cite{BEMS}
for the {\it same} partition function,
which looks more sophisticated and is very much in the spirit of \cite{Kle}
(that paper is in turn a development of \cite{Nek} and especially of \cite{EyKle}).
This representation is rather natural from the perspective of character
representations from \cite{Wops} and \cite{unint}, and is also closely related
to the matrix-model field theory of \cite{EyFT}
(a parallel development to \cite{AMM,Mth}).

The goal of this short note is to describe the explicit
relation between these two currently-available matrix model
representations of the simplest HK partition function:
one is nothing but the Fourier-Laplace transform of another.
This means that the pertinent advantages and relations of
each of these two models are immediately inherited by the
other, what can open additional possibilities for further
developments. The most important could be generalizations
to multiple and open Hurwitz numbers and a progress in
understanding of the puzzling Virasoro constraints and
related link between the HK partition function and conventional
cubic Kontsevich theory -- the main building block of the
M-theory of matrix models. See the last few papers in \cite{Hur}
for clearer formulation of these problems.

\bigskip

We want to find the shape of potential $V(M)$ from

\be
\boxed{
Z_{HK}\Big(\Psi = e^R\Big) = Z_{BEMS}(R),
}
\label{claim}
\ee
where

\begin{align}
Z_{HK}(\Psi) \ = \ & g^{-n^2/2} \exp\left(- g \dfrac{n(n^2-1)}{24} \right) \int
d \mu(\phi) \
\exp\left( - \dfrac{1}{2g} \tr \phi^2 +
\tr \big( e^{\phi - ng/2} \Psi \big) \right) = \nn \\ &
\nonumber \\ &
= g^{-n^2/2} \exp\left( - g \dfrac{n^3}{6} + g \dfrac{n}{24} \right) \int \dfrac{\Delta(\phi)}{\Delta(\Psi)} \prod\limits_{i = 1}^{n}
\exp\left( - \dfrac{\phi^2_i}{2g} - \dfrac{2n-1}{2} \phi_i + e^{\phi_i} \psi_i \right) d\phi_i
\label{zhk}
\end{align}
and

\begin{align}
Z_{BEMS}(R) \ = \ & g^{-n^2} \dfrac{\Delta( R )}{\Delta(e^{R})} \int dM \exp\left(-\dfrac{1}{g} \tr V(M) + \dfrac{1}{g} \tr MR \right)
= \label{zbemsx} \\ & \nonumber \\ & = g^{-n^2} \int \dfrac{\Delta(M)}{\Delta( e^{R} )}
\prod\limits_{i = 1}^{n} \exp\left( - \dfrac{m_i r_i - V(m_i)}{g} \right) dm_i
\label{zbems}
\end{align}
\smallskip\\
The integrals are taken over Hermitian $n \times n$ matrices $\phi, M$ and $R$ with eigenvalues $\phi_i, m_i$ and $r_i$, respectively. As usual, $\Delta(\phi) = \prod_{i < j} \big(\phi_i - \phi_j\big)$ and
$\Delta(e^\phi) = \prod_{i<j} (e^{\phi_i}-e^{\phi_j})$
are Van-der-Monde determinants. Note that, the $n$-dependent prefactors in the first and second lines of (\ref{zhk}) are different, because in the second line a convenient shift $\phi \mapsto \phi + ng/2$ is made and because of the exponential prefactor from (\ref{meas}), see \cite{Wops} for more details. Note also, that the $R$-dependent pre-factor in (\ref{zbemsx}) is important for
potential $V(M)$ to exist. According to (\ref{claim})-(\ref{zbems}) this
$V(M)$ is given by inverse Fourier-Laplace transform,

\be
e^{- \tr V(M)/g } = \int dR \dfrac{\Delta(e^R)}{\Delta(R)} \exp\left( -\dfrac{1}{g} \tr MR \right) Z_{HK} \big(e^R\big)
\ee
\smallskip\\
Note that $g^{-n^2}$ disappears after the transform. Substituting the expression for $Z_{HK}\big(e^R\big)$, we obtain

\begin{align*}
e^{- \tr V(M)/g } = g^{-n^2/2} \int dR \dfrac{\Delta(e^R)}{\Delta(R)} \exp\left( -\dfrac{1}{g} \tr MR \right) \int
d \mu(\phi) \ \exp\left( - \dfrac{1}{2g} \tr \phi^2 +
\tr \big( e^{\phi - ng/2} e^R \big) - g \dfrac{n(n^2-1)}{24}\right)
\end{align*}
The simplest option now is to diagonalize
$\phi = U \phi_{\rm diag} U^{-1},
R = {\widetilde U} R_{\rm diag} {\widetilde U}^{-1}$
and take Itzykson-Zuber integrals

\be
\int dU \exp (\tr UXU^\dagger Y) = \frac{\det_{ij} e^{x_iy_j}}{\Delta(X)\Delta(Y)}
\ee
\smallskip\\
over $U$ and ${\widetilde U}$. What remains is a pair of eigenvalue integrals:

\begin{align*}
e^{- \tr V(M)/g } = e^{ - g n^3/6 + gn/24} \int \prod\limits_{i = 1}^{n} \exp\left( - \dfrac{1}{g} m_i r_i \right) dr_i
\int \dfrac{g^{-n^2/2} \Delta(\phi)}{\Delta(M)}
\prod\limits_{i = 1}^{n} \exp\left( - \dfrac{\phi^2_i}{2g} - \dfrac{2n-1}{2} \phi_i + e^{\phi_i + r_i}
\right) d \phi_i
\end{align*}
Shifting $r_i \mapsto r_i - \phi_i$, we obtain

\begin{align*}
e^{- \tr V(M)/g } = e^{ - g n^3/6 + gn/24} \int \prod\limits_{i = 1}^{n} \exp\left( e^{r_i} - \dfrac{1}{g} m_i r_i \right) dr_i
\int \dfrac{ g^{-n^2/2} \Delta(\phi)}{\Delta(M)}
\prod\limits_{i = 1}^{n} \exp\left( - \dfrac{\phi^2_i}{2g} + \dfrac{1}{g} C_i \phi_i \right) d \phi_i
\end{align*}
where $C_i = m_i - g \dfrac{2n-1}{2} $. Using the Itzykson-Zuber formula and the Gaussian integral, it is easy to show that

\be
\int \Delta(\phi) \prod\limits_{i = 1}^{n} \exp\left( - \dfrac{\phi^2_i}{2g} + \dfrac{1}{g} C_i \phi_i \right) d \phi_i
= g^{n/2} \Delta(C) \prod\limits_{i = 1}^{n} \exp\left( \dfrac{C_i^2}{2g} \right)
\ee
Consequently,

\be
e^{- \tr V(M)/g } = g^{-n^2/2+n/2} \exp\left( - g\dfrac{n^3}{6} + g\dfrac{n}{24} + \dfrac{1}{2g} \sum\limits_{i = 1}^{n} \Big( m_i - g \dfrac{2n - 1}{2} \Big)^2 \right) \ \prod\limits_{i = 1}^{n} \int dr_i \exp\left( e^{r_i} - \dfrac{1}{g} m_i r_i \right)
\ee
\smallskip\\
The last integral can be taken explicitly by making a change of variable $\psi = e^r$:

\be
 f(m) = \int dr \exp\left( e^r - \dfrac{m}{g} r \right)
 = \int \dfrac{d\psi}{\psi} \exp\left( \psi - \dfrac{m}{g} \log \psi \right)
 = \int d\psi \ \psi^{- m/g - 1} e^{\psi} = e^{-i\pi m/g}\Gamma\big( - \dfrac{m}{g}\big)
\ee
\smallskip\\
The last equality is nothing but the integral representation of
(appropriate analytic continuation of) Euler's $\Gamma$-function.
It is, of course, correctly defined only for certain choice of integration contour.
We finally obtain

\begin{align}
e^{- \tr V(M)/g } = \prod\limits_{j = 1}^{n} e^{-V(m_j)/g} = \prod\limits_{j = 1}^{n} g^{(1-n)/2} \exp\left( - g\dfrac{n^2}{6} + \dfrac{g}{24} + \dfrac{1}{2g}
\Big( m_j - g \dfrac{2n - 1}{2} \Big)^2 - i \pi \dfrac{m_j}{g} \right)\Gamma\left(-\dfrac{m_j}{g}\right)
\end{align}
\smallskip\\
where the r.h.s. is indeed a product over all eigenvalues of $M$. Thus, the potential $V(M)$ exists and is equal to

\be
V(x) = i \pi x - g \log \Gamma\left(-\dfrac{x}{g}\right) + g^2\dfrac{n^2}{6} - \dfrac{g^2}{24} - \dfrac{1}{2}
\Big( x - g\dfrac{2n - 1}{2} \Big)^2 + \dfrac{n-1}{2} g \log g
\ee
\smallskip\\
or, after opening the brackets,

\be
\boxed{
V(x) = - \dfrac{x^2}{2} - g \log \Gamma\left(-\dfrac{x}{g}\right) + i \pi x + \left( n - \dfrac{1}{2} \right) g x + \dfrac{n-1}{2} g \log g  - \dfrac{g^2}{3} \left( n^2 - \dfrac{3}{2}n + \dfrac{1}{2} \right)
}
\label{result}
\ee
\smallskip\\
This expression coincides with eq.(8) of \cite{BEMS}. Note only that the very last term of this formula
in \cite{BEMS} contains $2$ instead of $1/2$, because of a misprint
in original expression for $f_\lambda(C_2)$ in that paper.
The right formula for $f_\lambda(C_2)$ is \linebreak
\be
f_\lambda(C_2) =
\frac{1}{2}\sum_{i=1}^n \lambda_i(\lambda_i-2i+1) =
\frac{1}{2}\sum_i h_i^2 - \left(n-\frac{1}{2}\right)\sum_i h_i
+ \frac{n}{3}\left(n^2-\frac{3}{2}n+\frac{1}{2}\right),
\ee
where $h_i = \lambda_i - i + n$, see eq.(26) in \cite{Wops}.
There is one more term in $V(x)$ in \cite{BEMS}, which is not present in
(\ref{result}): namely, $x \log g$.
It results from a slightly different definition of the HK partition function
in \cite{BEMS}:
it differs from that of \cite{Wops} by rescaling of the external field
$\Psi \rightarrow \Psi/g$.
This shifts $\ \tr MR = \tr M\log\Psi\ $ by $\ \log(1/g) \tr M$, what
exactly implies adding $x \log g$ to $V(x)$.

We conclude, that the two matrix models --
the model of \cite{BEMS} with the logarithm-of-a-gamma-function potential
and the model of \cite{Wops} with the trigonometric-Van-der-Monde determinant
-- are in fact equivalent.
Since this equivalence is now made explicit, both models can be used on equal footing.

\newpage

\section*{Acknowledgements}

Our work is partly supported by Russian Federal Nuclear Energy
Agency, by the joint grants 09-02-91005-ANF, 09-02-90493-Ukr,
09-01-92440-CE and 09-02-9305-CNRSL,
by the Russian President's Grants of
Support for the Scientific Schools NSh-3035.2008.2
and by the RFBR grant 07-02-00645.

\end{document}